\documentclass[reprint,superscriptaddress,amsmath,amssymb,aps,prr,floatfix]{revtex4-2}
\usepackage[usenames, dvipsnames]{color}

\usepackage{graphicx}
\usepackage{dcolumn}
\usepackage{bm}
\usepackage{siunitx}
\usepackage{hyperref}
\usepackage{nicefrac}
\usepackage{amsmath}
\renewcommand{\vec}{\boldsymbol}

\usepackage[capitalize]{cleveref}
\hypersetup{
    colorlinks,
    citecolor=blue,    filecolor=blue,
    linkcolor=blue,    urlcolor=blue
}
\usepackage{booktabs}

\begin{document}

\title{Pareto Optimization of a Laser Wakefield Accelerator}

\author{F. Irshad}
\author{C. Eberle}
\author{F.M. Foerster}
\author{K. v. Grafenstein}
\author{F. Haberstroh}
\author{E. Travac}
\author{N. Weisse}

\affiliation{Ludwig-Maximilian-Universit\"at M\"unchen, Am Coulombwall 1, 85748 Garching, Germany}
\author{ S. Karsch}

\author{A. D\"opp}

\affiliation{Ludwig-Maximilian-Universit\"at M\"unchen, Am Coulombwall 1, 85748 Garching, Germany} 
\affiliation{Max Planck Institut für Quantenoptik, Hans-Kopfermann-Strasse 1, Garching 85748, Germany}

\begin{abstract}

Optimization of accelerator performance parameters is limited by numerous trade-offs and finding the appropriate balance between optimization goals for an unknown system is challenging to achieve. Here we show that multi-objective Bayesian optimization can map the solution space of a laser wakefield accelerator in a very sample-efficient way. Using a Gaussian mixture model, we isolate contributions related to an electron bunch at a certain energy and we observe that there exists a wide range of Pareto-optimal solutions that trade beam energy versus charge at similar laser-to-beam efficiency. However, many applications such as light sources require particle beams at a certain target energy. Once such a constraint is introduced we observe a direct trade-off between energy spread and accelerator efficiency. We furthermore demonstrate how specific solutions can be exploited using \emph{a posteriori} scalarization of the objectives, thereby efficiently splitting the exploration and exploitation phases.
\end{abstract}

\maketitle

Laser-plasma acceleration (LPA) of electrons \cite{Esarey.2009,malka2012laser} and related radiation sources \cite{corde2013femtosecond} are an emerging technology with potentially broad applications in science, industry and medicine \cite{albert2016applications}. Over the past decade, there has been significant progress with regards to both quality and stability of the accelerated electron beams \cite{maier2020decoding} and recently, first experiments have demonstrated that beam parameters are sufficient to drive free electron lasers \cite{labat2022seeded,wang2021free}. Much of this progress can be attributed to improved performance of laser systems and better targets, in particular the use of controlled injection methods such as shock-front injection or ionization-induced injection \cite{clayton2010self, schmid2010density, wenz2019dual,gotzfried2020physics}. Nonetheless, the role of expert human operators in reaching optimal performance cannot be overstated. Their manual optimization typically relies on sequential line scans, using a combination of intuition and experience to determine optimal parameters for e.g. target position and laser pulse duration. However, the continuously increasing complexity of laser-plasma experiments has made reaching optimal performance in a reproducible manner increasingly difficult.

Machine learning (ML) techniques offer powerful tools to address this challenge. A particularly popular method for the optimization of laser-plasma accelerators is Bayesian optimization (BO) \cite{dopp2022data, Shalloo.2020, Jalas.2021}. BO is a global optimization method based on searching optima in a probabilistic surrogate model that is updated iteratively as the experiment progresses.  It is extremely sample efficient, meaning it converges to the optimum with relatively few measurement. This makes it very suitable for the optimization of laser-plasma accelerators, where measurements are often acquired at a relatively low acquisition rate. The underlying probabilistic model commonly referred to as a surrogate model is usually chosen as a Gaussian process (GP) \cite{williams1995gaussian, balandat2020botorch}. This is a non-parametric probabilistic model that assumes prior knowledge about possible relations between parameters and objectives. Within each iteration of the BO, a GP model is fitted to the current observations. This model is used to estimate a good position for the next measurement and the resulting observation is then appended to the data.
The objective of optimizing a laser-plasma accelerator typically consists of maximizing one or more metrics that are combined into a single scalar value using empirically determined weights \cite{Shalloo.2020, Jalas.2021}. In a previous publication \cite{Irshad.2023} we studied different objective functions such as combinations of mean electron energy, energy bandwidth and charge. There we found that \emph{a priori} definition of objective weights does often not lead to the desired outcome. Instead, we showed that Pareto optimization via expected improvement of the hypervolume occupied by all objectives in the output space can efficiently explore the trade-offs between objectives and be used to choose adequate Pareto-optimal solutions \emph{a posteriori}. The latter are all the best possible combinations of the objectives that are achievable, which form the so-called Pareto front.

\begin{figure*}[thp]
  \centering
  \includegraphics[width=1\linewidth]{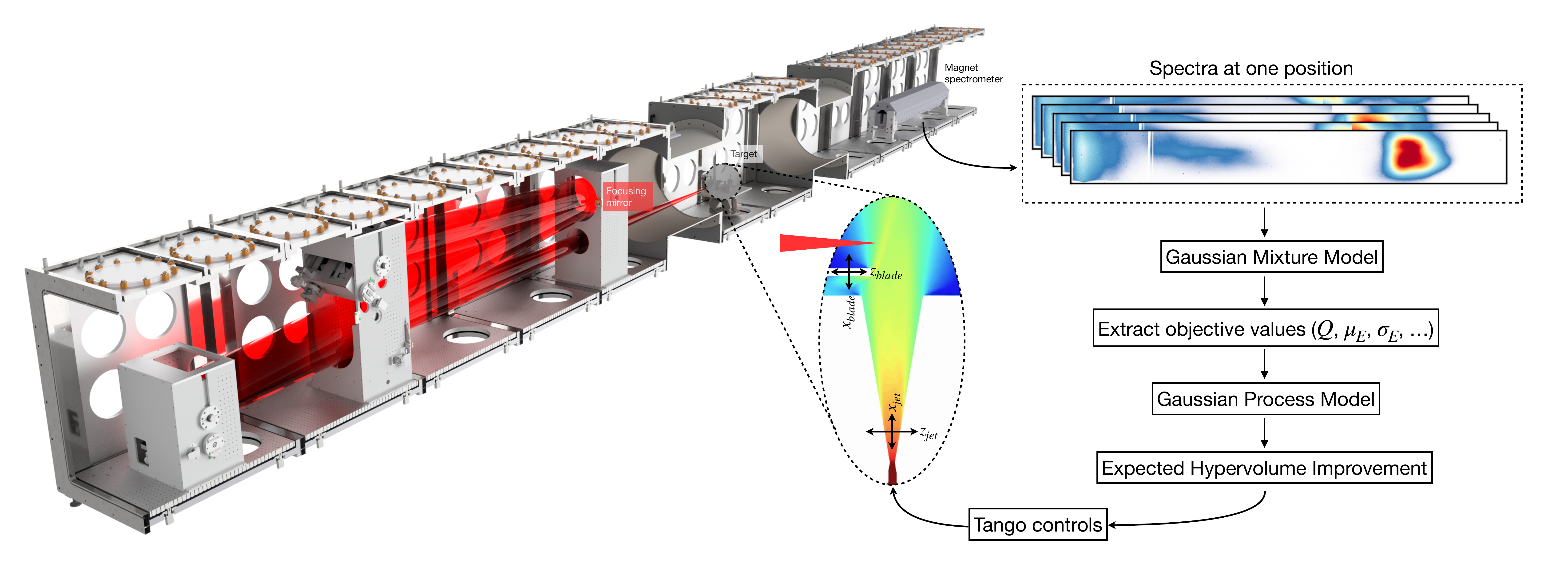}
  \caption{Sketch of the experimental setup and the optimization pipeline. The laser is focused in an f/44 geometry onto the target, which can be moved alongside a blade that locally perturbs the gas flow, shown as an inset with results from a fluid simulation of the target. The main diagnostic in this experiment is the dipole magnet spectrometer. At each setting investigated, $5-10$ shots are taken, cleaned using a Gaussian mixture model and the desired objective values are extracted. From there, a Gaussian process model is created that is probed to find candidate points with the highest expected hypervolume improvement. This optimal setting is then forwarded to Tango controls and a new set of data is taken. }
  \label{fig:experiment}
\end{figure*}

In this Letter, we present first experimental results on such a Pareto optimization of a laser wakefield accelerator. Experiments were performed with the ATLAS laser system at the Centre for Advanced Laser Applications in Garching, Germany. The experimental setup is illustrated in \cref{fig:experiment}. During the experiment the laser delivered $\SI{30} {fs}$ pulses at a central wavelength of $\SI{800} {nm}$ with an estimated energy of $ (4\pm 0.8)\:\si{\joule}$ energy within the $\SI{46}{\micro\meter}$ waist. The resulting peak intensity is $(7.5\pm0.4)\:10^{18}\:\si{\watt\per\cm^2}$, corresponding to a normalized vector potential of $a_0=1.8$.
The target consisted of a 7-mm-long supersonic gas jet using either pure Hydrogen or a mixture containing a nitrogen dopant. The jet is mounted on a hexapod for three-dimensional positioning. Moving the gas jet along the laser focus ($z$ axis) changes the peak intensity at the target entrance, while movement in the vertical ($x$) direction simultaneously changes the density and cross-section of the gas jet that the laser interacts with. A silicon wafer is mounted on an $(x,z)$ motorized stage to locally perturb the supersonic gas flow, leading to the formation of a shock front. At this shock front, the gas density rapidly drops, resulting in an expansion of the plasma wakefield that facilitates the injection of electrons \cite{schmid2010density}. The wafer can be moved along the gas jet to control the point of injection. This can be used to tune the energy of the electron beams since the acceleration length can be reduced or increased by moving the blade in or out of the gas jet, respectively. The blade is also motorized vertically over the gas jet to result in a different shapes of the shock \cite{guillaume2015control}. Tango controls is used for communicating with the different motors \cite{weisse2022tango}, allowing us to change the longitudinal gas jet position, gas jet height, blade position and blade height in an automated fashion \footnote{The scan ranges were chosen as follows: The gas jet was allowed to move within a range of 9 mm in $z$ and 4 mm in $x$. The blade in a range of 2.5mm in $z$ into the jet and 1.5 mm in $x$.}. A $\SI{80}{cm}$ long dipole magnet spectrometer with a magnetic field strength of $\SI{0.85}{T}$, situated $\SI{2.9}{m}$ downstream of the LWFA target, is used to disperse the electron beam onto several charge-calibrated scintillating screens \cite{Kurz2018}. Each screen is imaged using a CMOS camera and the measurements are stitched together to yield the complete electron energy spectrum.

\begin{figure}[t]
  \centering
  \includegraphics[width=.9\linewidth]{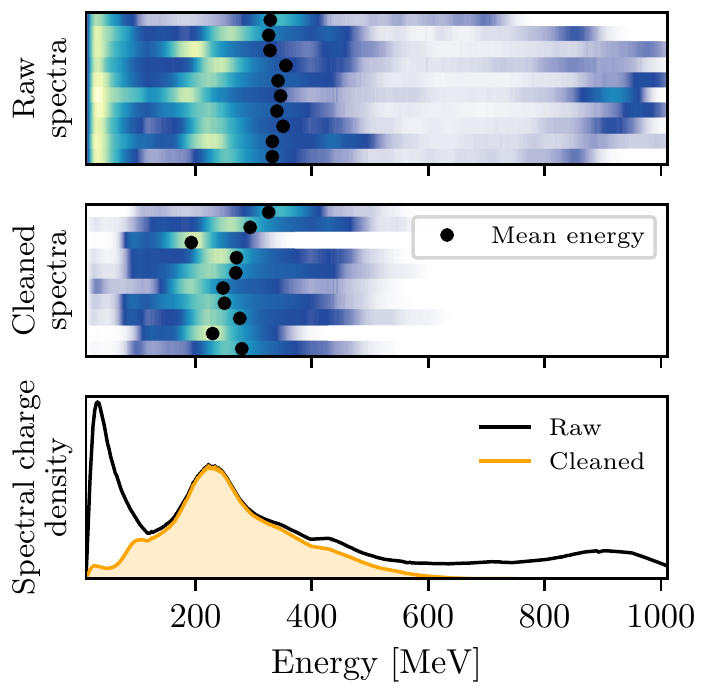}
  \caption{Measured raw electron spectra (top) and spectra after 'cleaning' using a Gaussian Mixture Model (center). The orange shaded line in the bottom plot shows the average cleaned spectrum versus the average original spectrum (black).}
  \label{fig:cleaned_spectra}
\end{figure}

From the energy spectrum we determine the total beam charge $Q$, mean energy $\overline{E}$ and energy spread $\sigma_E$. These parameters uniquely define a normally distributed energy spectrum and in their three-dimensional objective space $\mathcal{Y}$, their trade-offs span the Pareto "surface" $\mathcal{P}$. However, the spectral distribution of a laser-accelerated electron beam may considerably differ from a normal distribution and because of this, metrics such as the mean energy may not be characteristic of peaks in the spectral distribution. \cref{fig:cleaned_spectra} shows an example of such a multi-modal energy distribution. There a gas mixture is used and the resulting shock-assisted ionization injection \cite{thaury2015shock} is prone to produce low-energy tails in the spectrum, as well as a high-energy peak. While it would appear to the optimizer that this particular position in the input space $\mathcal X$ produced stable electron beams at $\SI{350}{\MeV}$, the spectrum is quite unstable, with most charge contained in a beam with fluctuating energy between $\SI{200}{\MeV}$ and $\SI{350}{\MeV}$. To separate the signal of individual bunches from the overall electron spectrum, we use a Gaussian mixture model (GMM)\cite{dopp2022data}. The GMM is an efficient and robust way to decompose the electron spectrum into its constituent parts, which can then be analyzed separately. As training data, we draw 1000 samples from a probability distribution derived from the electron spectrum. To initialize the GMM we use the number and position of local peaks within a 100 MeV range of the spectrum. We then use the Expectation-Maximization (EM) algorithm \cite{dempster1977maximum} to fit a multi-modal Gaussian distribution and divide the charge at each energy interval in the spectra according to its probability of belonging to a certain distribution. 
The effect of isolating bunches using GMM is illustrated in \cref{fig:cleaned_spectra}, showing that both the low and high energy contributions can be effectively removed. This postprocessing step takes on the order of tens of milliseconds and enabled us to isolate contributions related to bunches at certain target energies on the fly. This avoids contamination from other bunches and allows for a detailed analysis of each bunch. Thus, using the GMM we can more accurately calculate representative statistical measures of the target electron beam bunch. 

\begin{figure}[t]
  \centering
  \includegraphics[width=.9\linewidth]{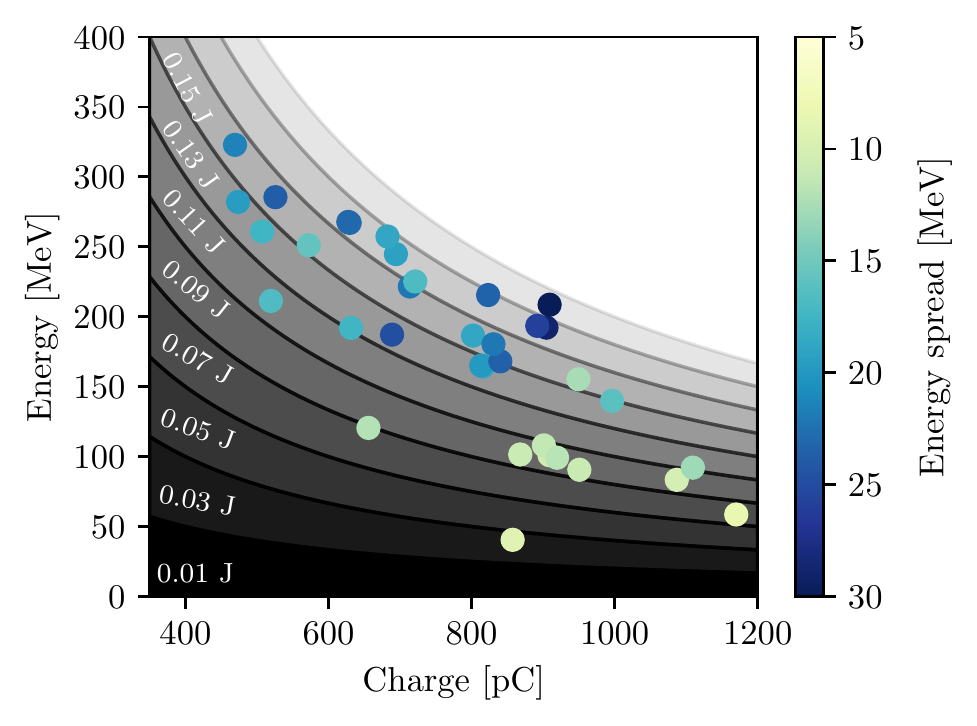}
  \caption{Pareto-optimal configurations of the laser wakefield accelerator for a Pareto optimization of beam energy, change and energy spread. The energy spread is color-coded. Regions of equal beam energy, and hence, accelerator efficiency, are shaded in grey. Note that the Pareto-optimal points form a surface and here we only show a projection of this space.}
  \label{fig:pareto1}
\end{figure} 

As a first optimization task, we maximize the beam charge within the bunch and its mean energy, while trying to minimize the energy spread. The optimizer takes $n=10$ shots $(\vec y_1(\vec x), \dots, \vec y_{n}(\vec x))$ at each position $\vec x$ in the parameter space $\mathcal{X}$ and calculates average values $\vec\mu(\vec x)=n^{-1}\sum_{i=1}^{n}\vec y_i(\vec x)$ to reduce the influence of shot-to-shot fluctuations. The total number of positions $\vec x$ taken in this optimization run was 60 including 5 random initial positions. The number of Pareto-optimal positions (also called non-dominated solutions \cite{dopp2022data}) resulting from this optimization run was 35, represented in \cref{fig:pareto1}. The product of beam charge and mean energy yields the total energy of the beam and is therefore a measure of energy transfer efficiency from the laser to the electron bunch. As such, there is an inherent trade-off between the two parameters. We observe that the Bayesian optimization can find this trade-off, as is visible in \cref{fig:pareto1}. Interestingly, we observe that most Pareto-optimal solutions cluster around lines of similar efficiency ($\sim 4\%$). Below 900 pC, high-charge solutions tend towards slightly higher efficiency, which is likely a result of higher beamloading leading to more energy extraction from the wakefield. However, as the charge approaches $\SI{1}{\nano\coulomb}$ we observe a substantial drop in efficiency, indicating a change in the operational regime. It is in this regime that we also observe beams with the lowest absolute energy spread. A physical interpretation of this behavior is that energy spread accumulates over acceleration and the only way to reach lower energy spread under our conditions is to also reduce the mean beam energy $\overline{E}(\vec x)$.
 
\begin{figure}[t]
  \centering
  \includegraphics[width=.9\linewidth]{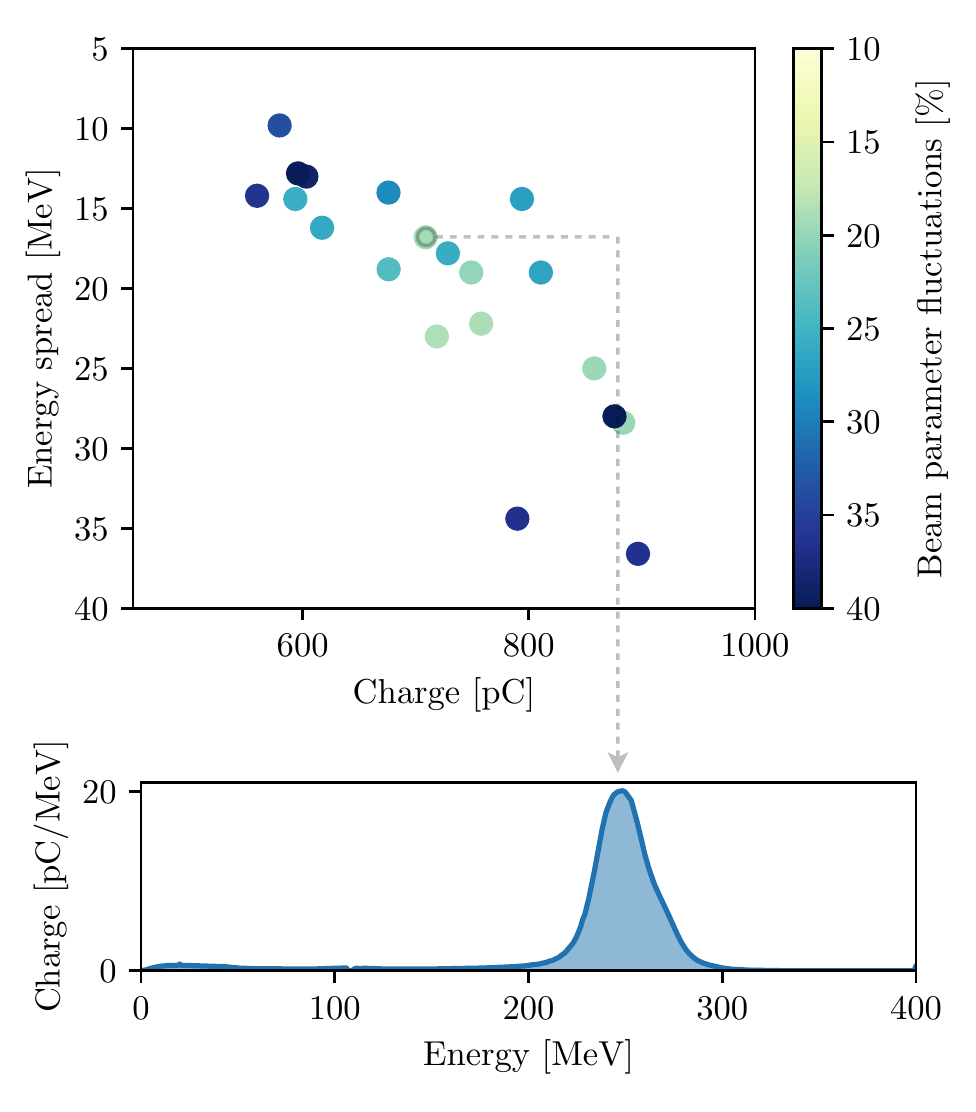}
  \caption{Pareto-optimization of a wakefield accelerator towards a target energy of $\SI{250}{\MeV}$. The top panel shows the non-dominated solutions, exhibiting a clear trade-off between charge and energy spread. The bottom shows the results of exploiting a single point on the Pareto front.}
  \label{fig:pareto2}
\end{figure}

 In the next step, we optimized for an electron beam at a certain target energy $E_{0}$. To achieve this, we introduce the objective $\Delta E(\vec x) = |E_{0}-\overline{E}(\vec x)|$, which is minimized the closer the expectation value of beam energy is to the target energy.  We also define an instability parameter $S(\vec x)$ in terms of the quadratic sum of relative fluctuations in each objective\footnote{This is calculated using the relative standard deviations $\sigma$ of the individual objectives, 
 \begin{equation}
S(\vec x) =  \sqrt{\left({\frac  {\sigma_{Q}(\vec x)}{\mu_Q(\vec x)}}\right)^{2}+\left({\frac{\sigma_{\sigma_E}({\vec x})}{\mu_{\sigma E}({\vec x})}}\right)^{2}+\left({\frac{\sigma_{\Delta E}({\vec x})}{\mu_{\Delta E}({\vec x})}}\right)^{2}} \ .
\end{equation}
This parameter can be interpreted as the sum of the standard deviations of the marginal distributions for $Q, \sigma_E$ and $\Delta E$ respectively. Therefore minimizing this parameter will also minimize the standard deviation of their joint distribution. We would like to note that this choice of the instability parameter can still be improved. A known issue with this definition is that stability naturally deteriorates as parameters such as the energy spread or distance to target energy are minimized.}. This instability parameter is used as a fourth objective dimension in the multi-objective optimization. The full objective vector is then given by $\vec y(\vec x)=(Q(\vec x), \sigma_E(\vec x), \Delta E(\vec x), S(\vec x))$. The effect of this is that the optimizer can find parameter combinations for which the electron beam is most stable with respect to the three objectives used. Given the explicit inclusion of stability objective into the model, we reduced the number of shots per position from 10 to 5 for this run. The total charge is maximized while the distance to target energy, beam bandwidth and instability parameter is minimized. The total number of positions taken in this optimization run was 65 (including the 5 random initial positions) and the Pareto-front is made up of 22 points.
The results for this optimization are shown in \cref{fig:pareto2}. As it is difficult to visualize the now four-dimensional objective space, we concentrate on the subspace of optimal charge, energy spread and stability. We observe a clear trade-off between charge and energy spread, an effect that has been observed in various papers related to beamloading \cite{Goetzfried.2020,Kirchen.2021,foerster2022stable}. Additionally, we can observe that there is often a choice between more stable settings with lower performance and less stable settings with higher performance.

As a global optimization method, there is no strict endpoint for Bayesian optimization \cite{lorenz2015stopping,nguyen2017regret}. The Pareto-optimization finds an estimate of the Pareto front $\mathcal{P}$ from the currently achievable parameters. But it will always try to improve upon this estimate by reducing uncertainty in its hypervolume-guided exploration of the parameter space. Thus, once the Pareto front is established to a sufficient degree \footnote{Here we use the simple criterion of a pre-defined number of iterations.} one should switch from exploration to exploitation of regions of the Pareto front that represent locally optimal solutions. To this end, we construct a single objective by choosing the hyper-parameters or weights of the individual objectives based on the generic form
 \begin{equation}
O_{\mathrm{sing}}(\vec x) = a\cdot Q(\vec x) + b\cdot \sigma_E(\vec x) + c\cdot \Delta E(\vec x),
\label{Single_Objective}
\end{equation}
where $a$, $b$ and $c$ are the scalar weights of each objective. Note that $a$ is restricted to be a positive real number, while $b$ and $c$ are restricted to be negative real numbers. The first step in the procedure of finding the best hyper-parameters of this single objective is to select a point ($\vec p\in\mathcal{P}\subset\mathcal{Y}$) in the region of the Pareto front that we want to exploit. The hyper-parameters should now be chosen such that they result in the highest value of the single objective for the chosen point. This problem can be reformulated into a loss function whose optimization yields the suitable hyper-parameters \footnote{For this purpose, we define the loss function
\begin{equation}
\ell(\vec p, \vec p') = \max_{\vec p'\in\mathcal{P}}\{ {O_{\mathrm{sing}}(\vec p')}\} - O_{\mathrm{sing}}(\vec p),
\end{equation}
which returns a value of zero when the chosen point has the maximum value of the single objective for a given set of hyper-parameters. This loss function is minimized to find the hyper-parameters ($a$, $b$, $c$) of the single objective function.}. These are then used to perform a single objective optimization. Here we use the lower-confidence bound acquisition function \cite{Irshad.2023}, which strictly exploits a region of the Pareto front and thereby converges to a local optimum. The result of such an optimization is shown at the bottom of \cref{fig:pareto2}, resulting in a beam of up to $\SI{20}{pC/MeV}$ in charge density at the target energy of $\SI{250}{\MeV}$.

To conclude, we have presented results on the Pareto optimization of a laser wakefield accelerator. Using a Gaussian mixture model, we were able to isolate contributions related to a bunch at a certain energy and we observed that there exists a wide range of Pareto-optimal solutions that trade beam energy versus charge with a laser-to-beam efficiency of around $5\%$. 
However, many applications such as light sources require particle beams at a certain target energy. Once such a constraint was introduced we observed a direct trade-off between energy spread and accelerator efficiency. We furthermore demonstrated how specific solutions can be exploited using an \emph{a posteriori} scalarization of the objectives, thereby efficiently splitting the exploration and exploitation phases. Importantly, our study has shown that Pareto-optimization is of interest from the point of optimization, allowing for an efficient choice of adequate trade-offs in an \emph{a priori} unknown system, but also regarding the physical interpretation of results. For instance, Pareto optimization is highly suitable to map the efficiency of a wakefield accelerator. One limitation of the present study is that the number of free parameters has been limited due to machine safety concerns. Studies in computer science have however shown that Bayesian optimization can be scaled to high-dimensional spaces \cite{Wang.2013,Wesley.2021} and we plan to include more degrees of freedom such as spectral phase and wavefront in future studies.

The findings of this study indicate that Pareto optimization as a method for experiment optimization and result interpretation can have a significant impact on the way experiments are conducted, not restricted to laser-plasma acceleration, but also as a general optimization method in physics research. Furthermore, the proposed procedure to separate the multi-objective exploration and single-objective exploitation phases provides a clear methodology for efficiently separating these two phases of optimization that can be translated to many other use cases.


%

\end{document}